\documentclass[runningheads]{llncs}
\pdfoutput=1
 
\usepackage{eccv}

\usepackage{eccvabbrv}

\usepackage{graphicx}
\usepackage{booktabs}

\usepackage[accsupp]{axessibility}  

\usepackage[pagebackref,breaklinks,colorlinks,citecolor=eccvblue]{hyperref}

\usepackage{orcidlink}

\begin{document}

\title{FastTalker: Jointly Generating Speech and Conversational Gestures from Text} 

\titlerunning{FastTalker}

\author{Zixin Guo\inst{1},
Jian Zhang\inst{2}}

\authorrunning{Zixin Guo and Jian Zhang}

\institute{
University of Toronto\\
\email{zixin.guo@mail.utoronto.ca}\and 
Waseda University\\
\email{zjlando@asagi.waseda.jp} 
}

\maketitle

\begin{abstract}
Generating 3D human gestures and speech from a text script is critical for creating realistic talking avatars. One solution is to leverage separate pipelines for text-to-speech (TTS) and speech-to-gesture (STG), but this approach suffers from poor alignment of speech and gestures and slow inference times. In this paper, we introduce FastTalker, an efficient and effective framework that simultaneously generates high-quality speech audio and 3D human gestures at high inference speeds. Our key insight is reusing the intermediate features from speech synthesis for gesture generation, as these features contain more precise rhythmic information than features re-extracted from generated speech. Specifically, 1) we propose an end-to-end framework that concurrently generates speech waveforms and full-body gestures, using intermediate speech features such as pitch, onset, energy, and duration directly for gesture decoding; 2) we redesign the causal network architecture to eliminate dependencies on future inputs for real applications; 3) we employ Reinforcement Learning-based Neural Architecture Search (NAS) to enhance both performance and inference speed by optimizing our network architecture. Experimental results on the BEAT2 dataset demonstrate that FastTalker achieves state-of-the-art performance in both speech synthesis and gesture generation, processing speech and gestures in 0.17 seconds per second on an NVIDIA 3090. 

\end{abstract}

\begin{figure}[tb]
  \centering
  \includegraphics[height=9cm]{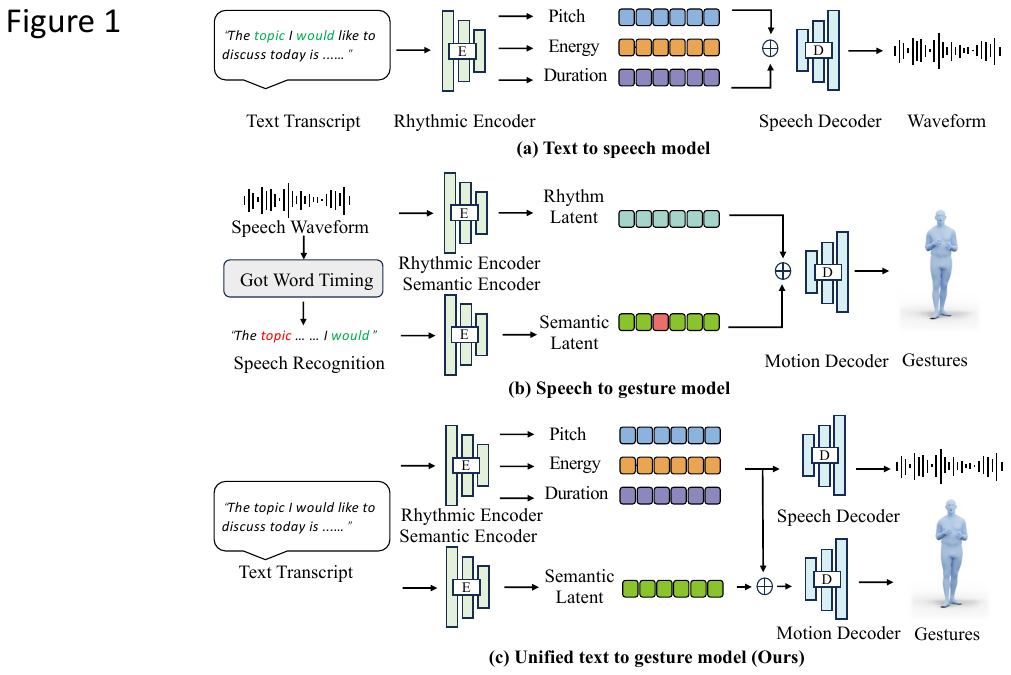}
\caption{\textbf{Conceptual Comparison}. Previous works process (a) text to speech and (b) speech to gesture separately, generating gestures with redundant encoders and inaccurate latent features. We propose a unified model (c) that reuses intermediate features to jointly generate speech and gestures efficiently and effectively.}
  \label{fig:1}
\end{figure}

\section{Introduction}
\label{sec:intro}
The automated generation of speech audio and full-body gestures from text scripts for daily conversations, like "\textit{the first thing I would like to do on weekend is hiking...}", has growing interest in both academic and industry due to its vast potential applications in talking avatars for games and virtual live streaming. 

Close to this task, existing works have separately discussed text-to-speech (TTS) \cite{ren2019fastspeech, ren2020fastspeech, shen2023naturalspeech, ju2024naturalspeech, tan2024naturalspeech, kim2021conditional}, speech-to-gesture (STG) \cite{richard2021meshtalk, fan2022faceformer, xing2023codetalker, yoon2020speech, liu2022learning, liu2022beat, yi2023generating, liu2023emage} and text-to-gesture (TTG) \cite{bhattacharya2021text2gestures}. However, to the best of our knowledge, there is no open-source work that jointly generates speech and gestures from textual scripts. The most straightforward solution is combining state-of-the-art methods such as FastSpeech 2 \cite{ren2020fastspeech} and EMAGE \cite{liu2023emage} for TTS and STG separately. 

However, straightforward separated pipeline is neither efficient nor effective. As shown in Fig. \ref{fig:1}, the (a) text-to-speech pipeline \cite{ren2020fastspeech} generates speech with estimated speech gestures intermediate features, \textit{i.e.}, the pitch and energy of speech. (b) speech-to-gesture model \cite{liu2023emage} generates gestures by estimated speech rhythmic features and speech semantic features from predicted text scripts. Our key insight is: directly leveraging the intermediate features from the TTS pipeline can enhance the accuracy of speech-audio alignment while eliminating the redundant speech encoders and word timing model in the STG pipeline. Therefore, we propose a pipeline (c) that directly leverages speech synthesis features—specifically pitch, energy, and duration—as inputs for the gesture decoder. 

The proposed approach, jointly generating speech and gestures, has the potential to be a efficient end-to-end framework. However, the potential for fast inference is constrained by the following factors: i) The bi-directional design of the original Transformer-based architecture \cite{devlin2018bert, brown2020language} in FastSpeech2 and EMAGE requires inputs from future text to optimize performance. ii) The high computational demand of the deep networks and transformer leads that generating two seconds of motion and speech at 15 FPS takes an average of 5.2 seconds on an NVIDIA 3090 GPU, which is impractical for real-time applications. 

To address these, we revisit the architecture in FastSpeech2 and EMAGE, and re-implement all self and cross-attention with casual mode and design the network to be shallow, e.g., around eight layers. But, i) adopting uni-directional information, ii) leveraging one rhythm latent for both speech and gesture generation and iii) decrease the networks complexity resulted in the drop of performance. To improve this, we employ a search-based method to optimize the proposed network's architecture hyper-parameters, such as the number of layers in encoders and decoders, kernel sizes, and channel dimensions. Specifically, we utilize Reinforcement Learning-Based Neural Architecture Search (NAS). The optimization process shows that surprisingly simpler MLP-based decoder for gestures, performs on bar with state-of-the-art TTS models and outperform existing STG models in both quality and speed. Our contributions can be summarized as follows:
\begin{itemize}
    \item We explore the task of jointly generating speech and co-speech gestures from text scripts, beginning with a baseline that separately combines state-of-the-art TTS and STG methods.
    \item We present FastTalker, an end-to-end framework designed to enhance speech-gesture alignment and improve the efficiency of baseline model.
    \item We significantly accelerate the inference speed of FastTalker by introducing a causal network architecture and refining the performance through neural architecture search.    
\end{itemize}

\section{Related Work}
\subsection{Text to Speech Synthesis}
Recent advances \cite{peng2020non,huang2019voice,li2019neural,jia2018transfer,ren2019fastspeech,ren2020fastspeech,li2023stock,li2024feature,zhang2024research,lyu2022study,lyu2023attention} in TTS synthesis can be categorized into three main groups: The FastSpeech series, FastSpeech \cite{ren2019fastspeech} and FastSpeech2 \cite{ren2020fastspeech}, exemplifies the efficiency of transformer-based architectures in speech synthesis, optimizing computational cost by decoupling duration prediction from acoustic feature generation. The NaturalSpeech series \cite{shen2023naturalspeech, ju2024naturalspeech, tan2024naturalspeech}, offers significant advancements in producing natural-sounding speech based on improved prosody and expression. But they are non-open-source for the research community. Finally, Flow-based Methods, VITS \cite{kim2021conditional} and VITS 2 \cite{kong2023vits2}, stand out as a conditional variational autoencoder with a flow-based generative model. While it provides high-quality speech synthesis and the flexibility of a non-autoregressive model. But its integration with other pipelines can be challenging as the inversed training procedures for flow-based model. Given its balance of performance, ease of use, and adaptability in multi-modal tasks, we select FastSpeech 2 as our baseline.

\subsection{Speech to Gesture Generation}
Existing approaches in speech-to-gesture synthesis typically fall into two groups: separately modeling facial and body animations, and uniform full-body generation. In the separate modeling category, facial models \cite{richard2021meshtalk, fan2022faceformer, xing2023codetalker,lu2023artistic,danvevcek2023emotional, peng2023emotalk, cudeiro2019capture} such as MeshTalk \cite{richard2021meshtalk}, synthesize 3D facial expressions directly from speech. FaceFormer \cite{fan2022faceformer} introduces a transformer-based model that generates dynamic facial movements, while CodeTalker \cite{xing2023codetalker} extends these capabilities by integrating quantization with pretrained VQVAE. For body animation \cite{ginosar2019learning, yoon2020speech, liu2022learning, liu2022beat, liu2022disco, pang2023bodyformer, li2021audio2gestures, ahuja2020style}, Trimodal \cite{yoon2020speech} leverages speech and text to animate upper body gestures. HA2G \cite{liu2022learning} introduces a hierarchical attention model that maps acoustic features to gestures, while CaMN \cite{liu2022beat} leverages a cascaded network for refining body and hand gestures. BodyFormer \cite{pang2023bodyformer} attempts to synthesize body animations from speech by introducing a novel transformer-based framework. Mix-Stage \cite{ahuja2020style} achieves style transfer for co-speech gesture generation by learning unique style embeddings for each speaker. On the other hand, full-body approaches \cite{habibie2021learning, yi2023generating, liu2023emage} attempt to cohesively generate both facial and bodily gestures for avatar animation. Talkshow \cite{yi2023generating} first introduces the VQVAE in gesture generation to improve diversity, and EMAGE \cite{liu2023emage} extends VQVAEs with lower body and global motion animations, introducing a new SMPLX \cite{SMPL-X:2019} and FLAME \cite{FLAME:SiggraphAsia2017} based gesture generation benchmark, BEAT2. We select EMAGE as our baseline as it achieves state-of-the-art performance by concurrently synthesizing facial expressions and body gestures and is implemented in straightforward transformer-based networks.

\subsection{Text to Gesture Generation}
Most similar to ours, Text2Gesture \cite{bhattacharya2021text2gestures} utilizes a transformer-based auto-regressive decoder to synthesize gestures from word sequences without speech in the middle stage. However, unlike speech, which naturally embodies temporal dynamics and prosodic cues, text alone provides limited direct information about the timing or intensity of gestures. This limitation primarily affects the accuracy of estimating gesture durations from text scripts and leads to animations that feel robotic or out of sync with spoken content.

\subsection{Neural Architecture Search}
Neural Architecture Search (NAS) \cite{xie2021weight,hua2024v2xum,hua2024finematch,yu2024promptfix,tang2024avicuna,yu2022cyclic, yang2021towards, dong2024automated, chen2020stabilizing,liu2020reinforcement,li2023zico, ye2022b} has been proposed to automatically search for high-performance networks and can be divided into RL-based and differentiable NAS. RL-based NAS methods utilize the REINFORCE algorithm to optimize the parameters of the controller, which is a recurrent neural network (RNN). The parameters of the controller determine the sampling policy for generating a sequence of symbols that specify the architecture of the child architecture. The reward function is based on the designed test accuracy of child architectures. On the other hand, differentiable NAS, such as DARTS \cite{liu2018darts}, introduces a differentiable search algorithm, which relaxes the search space to be continuous. However, compared to RL-based methods, the robustness of differentiable-based methods is limited, e.g., the collapse of DARTS. Therefore, we adopt RL-based NAS, which is more robust in real applications, for improving the performance of speech and gesture generation.

\begin{figure}[tb]
  \centering
  \includegraphics[height=6cm]{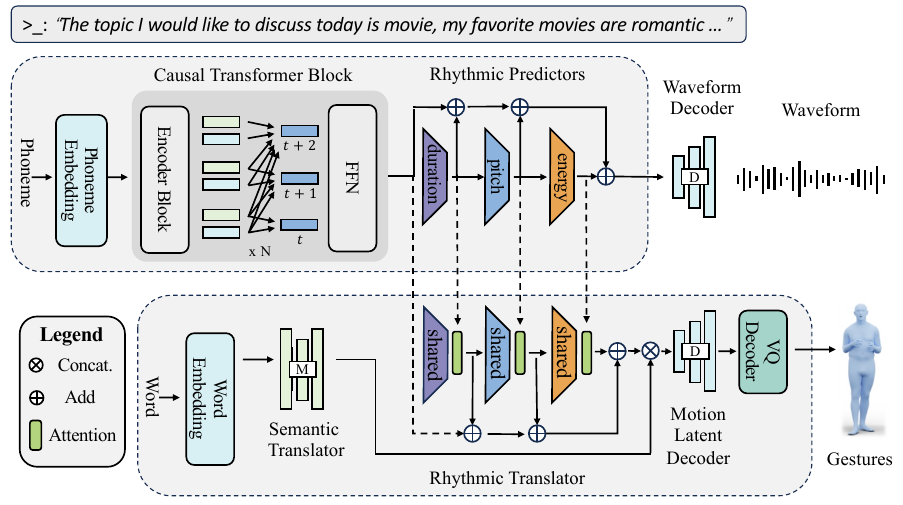}
  \caption{\textbf{FastTalker} first preprocesses text scripts into phoneme and word sequences. For speech generation (top), initial features are encoded by a causal transformer encoder, and then separate rhythm predictors are adopted to estimate intermediate speech duration, pitch, and energy to decode speech. For gesture generation (bottom), it employs shared rhythm predictors to predict gesture rhythm features and fuse them with the speech features via attention. The merged features are concatenated with semantic features extracted from word embeddings for gesture latent decoding. Finally, gestures are reconstructed from the latent features by pretrained VQVAEs.}

  \label{fig:2}
\end{figure}

\section{FastTalker}
\label{sec:3}
In this section, we introduce FastTalker, as shown in Fig. \ref{fig:2}. FastTalker takes a text script as input and outputs synchronized speech and gesture sequences through two branches. Initially, the speech generation branch encodes the text into phoneme-level embeddings to extract intermediate speech rhythmic features and utilizes task-specific decoders for generating explicit speech rhythm features, as described in Section \ref{sec:31}. Then, the gesture generation branch employs shared rhythm predictors to fuse the rhythmic features to generate gesture-specific rhythm features, and then combines semantic features from the text to decode gestures, as detailed in Section \ref{sec:32}.

\subsection{Speech Generation Branch}
\label{sec:31}
FastTalker begins by extracting a joint rhythmic latent representation for speech and gesture generation from sentence-level input text scripts, such as "\textit{The topic I would like to discuss today is...}". Following the preprocessing approach of FastSpeech2, we convert the text into its phonemic components, for example, "\textit{th, e, to, ...}". For a script containing \(k\) words, we derive \(n\) phonemes, initializing the phoneme embedding \(e_{pho}\), where \(e_{pho} \in \mathbb{R}^{n \times l}\) and \(l\) represents the feature length of each phoneme embedding.

The phoneme embedding is refined by a Causal Transformer Encoder, as shown in Fig. \ref{fig:2}. It utilizes feed-forward transformer encoder blocks for temporal feature aggregation from the learnable phoneme embeddings. Unlike FastSpeech2, our encoder employs a causal self-attention mechanism to facilitate real-time applications. The temporally encoded phoneme features \(f_{pho}\) are then mapped to explicit rhythmic features, including pitch \(p\), duration \(d\), and energy \(e\), using separate residual-based predictors. 
As defined in FastSpeech2, the phoneme duration \(d \in \mathbb{R}^n\) indicates the duration of speech sounds, with the corresponding total speech length given by \(m = \sum(d)\). Pitch \(p \in \mathbb{R}^m\) represents speech prosody and conveys emotions, while energy \(e \in \mathbb{R}^m\) influences the volume and prosody of speech, representing the frame-level intensity of the mel-spectrograms.

The architecture of these predictors is shown in Fig. \ref{fig:2}. All the duration, pitch, and energy predictors consist of the same CNN-based network blocks, comprising a two-layer 1D convolutional network with ReLU activation. The configuration of these networks, including feature length and depth, is optimized through Neural Architecture Search (NAS) as discussed in Section \ref{sec:4}. Each block also incorporates dropout and layer normalization. The final block has an additional linear layer for mapping the latent features to duration, pitch, and energy ground truth values.

The ground truths for duration, pitch, and energy are pre-calculated by preprocessing to train FastTalker in a supervised manner. The duration is determined using Montreal Forced Alignment (MFA), indicating the number of audio frames corresponding to each phoneme. We convert this into the logarithmic domain and optimize it using mean square error (MSE) with the output from the duration predictor \(P_d\) as 
\begin{equation}
    L_{duration} = \| \log(d) - M_d(P_d(f_{pho})) \|_2^2.
\end{equation} 
where \(M_d\) is the linear projection head for duration. 

The pitch predictor \(P_p\) leverages features from before the linear projection head, \(f_d\) and the original \(f_{pho}\), to estimate pitch. The pitch ground truth is derived from a Continuous Wavelet Transform (CWT), decomposing the continuous pitch series into a pitch spectrogram. Pitch training also employs MSE loss as 
\begin{equation}
    L_{pitch} = \| p - M_p(P_p(f_d + f_{pho})) \|_2^2.
\end{equation}
Lastly, the energy value is calculated by taking the L2-norm of the amplitude for each short-time Fourier transform (STFT) frame, with energy prediction optimized using MSE loss between the output of the energy predictor and the ground truth energy values as 
\begin{equation}
    L_{energy} = \| e - M_e(P_e(f_d + f_{pitch} + f_{pho})) \|_2^2.
\end{equation}
After these, we obtain the explicit rhythmic features \(f_r = \{f_d, f_{pitch}, f_e\}\) using separate predictors for the decoding of speech audio and motion sequences.

For speech waveform decoding, similar to FastSpeech2, we utilize the mel-spectrogram \(s \in \mathbb{R}^{80 \times m}\) as explicit speech features to enhance the training of the speech waveform decoder \(D_{wave}\). The mel-spectrogram decoder \(D_{mel}\) is trained concurrently with the waveform decoder using MSE loss. However, during inference, we omit \(D_{mel}\) and solely utilize \(D_{wave}\). The speech waveform decoder incorporates dilated convolutions, gated activations, and Conv1D layers to effectively capture both short and long-term audio dependencies. The configuration of this network, including the number of layers, is also determined through Neural Architecture Search (NAS) as detailed in Section \ref{sec:4}. This decoder processes the concatenated features \(f_r\) and employs transposed 1D-convolution to align with the length of the audio clip \(a\). The training of the speech waveform decoder and mel-spectrogram decoder is supervised using a combination of multi-resolution STFT loss, LSGAN discriminator loss, and MSE loss. The total loss \(L_{audio}\) is computed as follows:
\begin{equation}
    L_{audio} = \text{STFT}(a, D_{wave}(f_r)) + \text{Adv}(a, D_{wave}(f_r)) + \text{MSE}(s, D_{mel}(f_r))
\end{equation}

\subsection{Gesture Generation Branch}
\label{sec:32}
Our key insight is to reuse the intermediate rhythmic features \(f_r = \{f_d, f_{pitch}, f_e\}\) and share the weights for their predictors to simultaneously generate speech and gestures. Unlike speech, gestures respond differently to the pitch and energy of speech and are only weakly correlated. To bridge this gap, we introduce a feature fusion approach.
As shown in Fig. \ref{fig:2}, the rhythmic translator \(T_r\) consists of shared rhythm predictors and additional attention layers. It encodes the refined phoneme features \(f_{pho}\) into gesture-specific rhythmic features \(r_g\) and fuses the reference rhythm features, e.g., duration \(f_d\), pitch \(f_{pitch}\), and energy \(f_e\), after each predictor with attention. The attention layer is a Temporal Convolutional Network (TCN) that generates weights from zero to one for weighted sum of the reference and gesture-specific features. This design enables us to adopt only three shallow CNN-based attention layers to obtain gesture rhythm features.  

In addition to rhythm features, as in EMAGE, we utilize pretrained word embeddings as the semantic features for gesture generation. The semantic translator \(T_s\) combines a 300-dimensional embedding from FastText with a Multi-Layer Perceptron (MLP) to adjust the dimensions suitable for the gesture decoder. Unlike co-speech gesture generation, which often uses words estimated by Auto Speech Recognition (ASR), our framework adopts the ground truth words to ensure the accuracy of the semantic features \(s_g\).

The concatenated semantic latent \(s_g\) and rhythmic latent \(r_g\) are then input into the gesture latent decoder \(D_l\), which maps the rhythmic and semantic features from text to the latent codes of gestures. Following the approach of EMAGE, the gesture latent codes \(l_g\) are from pretrained motion Vector Quantized Variational Autoencoders (VQVAEs), where separate VQVAEs are used for the face, upper body, lower body, and hands. These VQVAEs encode the gestures' rotation representations in a codebook \(c \in \mathbb{R}^{256 \times 256}\), where 256 is the number of codebook indices and the feature length. This quantization of continuous gesture features into discrete codes transforms the regression loss into a classification loss, significantly enhancing gesture diversity \cite{talkshow:yi2022generating}. Therefore, the training of the gesture latent decoder \(D_l\) is optimized by both codebook index classification loss and latent reconstruction loss as:
\begin{equation}
     L_{gesture} = E(l_g, \hat{l}_g) + \|l_g - \hat{l}_g\|_1,
\end{equation}
where \( E \) denotes the cross-entropy function.

Once the gesture latent codes are obtained, we utilize the pretrained Gesture VQVAE Decoders \(D_g\) and Global Motion Decoder \(D_m\) to reconstruct the gestures in rotation space. Since these VQVAE decoders consist of shallow convolutional layers, they are computationally efficient. We fix their pretrained weights from EMAGE during training and inference.

\begin{figure}[tb]
  \centering
  \includegraphics[height=6cm]{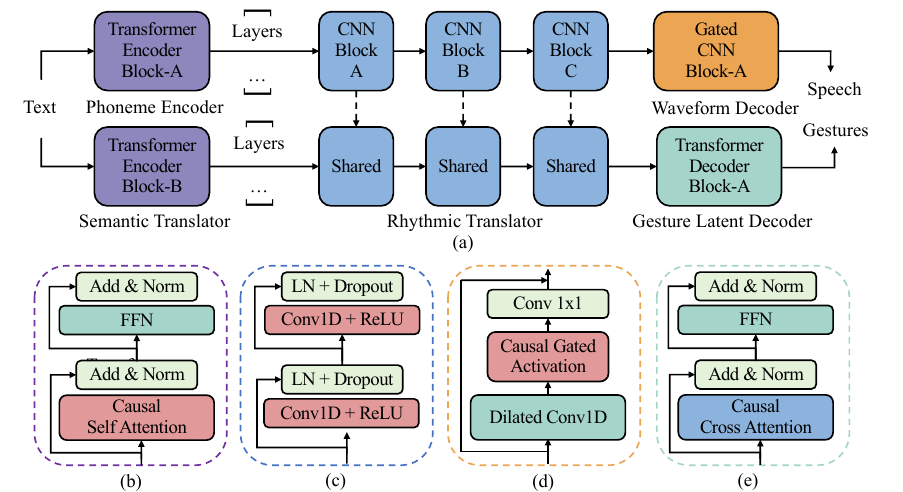}
\caption{\textbf{Neural Architecture Search for FastTalker}. (a) The architecture of FastTalker is decomposed into four types of basic network blocks including: Transformer Encoder Block (b), CNN Block (c), Gated CNN Block (d), and Transformer Decoder Block (e) for the phoneme/semantic encoder, rhythm translator, speech decoder, and gesture decoder respectively. The hyperparameters of these blocks, such as the number of layers, are optimized with NAS.}

  \label{fig:3}
\end{figure}

\section{Neural Architecture Search for FastTalker}
\label{sec:4}
\subsection{Search Space of FastTalker}
A key motivation for utilizing unified intermediate rhythmic features and rhythm predictors for speech and gesture generation in FastTalker is to enhance inference speed for real-time applications. We initially describe the basic architecture of FastTalker in Section \ref{sec:3}. The design of causal attention-based transformers and shared rhythmic feature predictors reduces the computation complexity but slightly decreases the generation performance. 
Moreover, the deep Transformer and convolution layers used in the original FastSpeech2 and EMAGE architectures, typically comprising over 40 layers, lead to inference speeds of approximately 4 seconds for generating 2 seconds of speech and motion. These challenges require a shallower and stronger network architecture for FastTalker. Reducing the number of layers usually presents a trade-off between generation performance and inference speed. However, we consider that in FastTalker, the rhythm components such as pitch, energy, and duration vary in training complexities. The predictors for estimating each component should have different optimal hyperparameters (number of layers, etc.). Finding the best hyperparameters, as shown in Fig. \ref{fig:3}, for each module like the predictor, has the potential to boost performance while reducing complexity. 

However, manually searching for the hyperparameters is challenging due to the size of the search space. Specifically, the module-level search space includes seven modules: phoneme encoder, duration predictor, energy predictor, pitch predictor, semantic translator, waveform decoder, and gesture latent decoder. For each module, the search involves combinations of feature channels \(32, 64, 128, 256\), layers \(0, 2, 4, 8\), and the number of group convolutions \(1, 2, 4, 8\). This results in a vast search space of \(7^{4 \times 4 \times 4}\), making it impractical to optimize through grid search. Thus, we implement Reinforcement Learning (RL) based Neural Architecture Search (NAS) to identify the optimal network architecture for this specific task of generating speech and gestures. 

In particular, a policy network acting as a controller is trained to sample candidate neural networks by estimating the probability of each hyperparameter configuration. We sample candidate networks based on the configuration. Each candidate is then trained to convergence and evaluated on performance metrics to provide feedback rewards to the controller. The controller, based on a Recurrent Neural Network (RNN), is trained using Reinforcement Learning (RL) due to the non-differentiable nature of the rewards relative to the network hyperparameters. The architecture for FastTalker is defined through a list in \(7 \times 4\), containing the possibilities for selecting specific hyperparameters. For example, the first output of the RNN is softmaxed into four class possibilities \(0, 0.9, 0.1, 0\), representing the selection of channel 64 in the feature channel search space \(32, 64, 128, 256\).

\subsection{Training with Reinforcement Learning}
The RNN controller is trained using the policy gradient algorithm \cite{sutton2000policy} with a length of 24 to accommodate four hyperparameters per module across seven modules. We sample a candidate architecture of FastTalker by maximizing the output probabilities of the RNN. The rewards for the controller are computed based on several metrics: FGD \cite{yoon2020speech}, UTMOS of speech \cite{saeki2022utmos}, and the number of total network parameters. These performance metrics are normalized for balanced optimization across gesture, speech, and inference speed evaluations. The training of each candidate network continues with a fixed-stop policy, where all candidate networks are trained for 300 epochs. Mathematically, the controller RNN outputs a sequence of hyperparameters \(h_{1:T}\) and receives a reward signal \(R\) calculated as
{\small
\begin{equation}
    R = \frac{\alpha}{\text{FGD}} + \text{UTMOS} + \frac{\beta}{\text{parameters}},
\end{equation} 
}
where \(\alpha\) and \(\beta\) are normalization factors. Training of the controller employs the REINFORCE algorithm, which adjusts the probability of selecting specific hyperparameters based on reward values, favoring hyperparameters that are likely to yield higher rewards. The optimization objective \(J(\theta_c)\) is defined as:
{\small
\begin{equation}
J(\theta_c) = \sum_{t=1}^{T}\left(d(h_{(t-1):1})\sum_{h_t \in H}P(h_t|h_{(t-1):1}; \theta_c)R\right)
\end{equation}
}
where \(\theta_c\) denotes the controller parameters, \(d(x)\) is the probability distribution under the policy \(P(h_t|a_{(t-1):1}; \theta_c)\), and \(H\) represents the hyperparameter space. Training directly optimizes the parameterized stochastic policy \(P(h_{1:T}; \theta_c)\) through gradient ascent on the reward function. The gradient updates for \(\theta_c\) are based on the non-differentiable reward signal \(R\) using the REINFORCE rule:
{\small
\begin{equation}
\nabla_{\theta_c}J(\theta_c) = \sum_{t=1}^{T}\left(d(h_{(t-1):1})\sum_{h_t \in H}\nabla_{\theta_c}P(h_t|h_{(t-1):1}; \theta_c)R\right)
\end{equation}
}
In our setup, the gradients \(\nabla_{\theta_c}P(h_t|h_{(t-1):1})\) are known, allowing us to define the likelihood-to-ratio function \(\nabla_{\theta_c}D(h_t|h_{(t-1):1})\) as the score function. To simplify and stabilize the training process, we use the expected value under the policy:
{\small
\begin{equation}
\nabla_{\theta_c}J(\theta_c) = \sum_{i=1}^{T}E_{P(h_{1:T}; \theta_c)}\left[\nabla_{\theta_c}\log P(h_t|a_{(t-1):1}; \theta_c)R\right]
\end{equation}
}
An empirical approximation of this function is:
{\small
\begin{equation}
\frac{1}{o}\sum_{k=1}^{o}\sum_{t=1}^{T}\nabla_{\theta_c}\log P(h_t|h_{(t-1):1}; \theta_c)R_k
\end{equation}
}
where \(T\) is 24, representing the number of hyperparameters predicted by the controller, and \(o\) is the number of different architectures sampled in one batch. To reduce the variance in estimation, we use a baseline function \(b\), computed as an exponential average of previous rewards:
{\small
\begin{equation}
\frac{1}{o}\sum_{k=1}^{o}\sum_{t=1}^{T}\nabla_{\theta_c}\log P(h_t|h_{(t-1):1}; \theta_c)(R_k - b)
\end{equation}
}
This baseline helps ascertain whether the current reward is significant compared to previous outcomes, enhancing the stability of the controller's training.

\section{Experiments}
\subsection{Settings}
\subsubsection{Dataset.} Our experiments utilize the BEAT2-Standard dataset \cite{liu2023emage}, which comprises 26 hours of synchronized text scripts, speech, and motion data captured via motion capture (mocap) recording. The dataset features actors engaging in daily conversations on topics such as hobbies, with accompanying body gestures. Speech data is provided in raw waveform format at a sampling rate of 16 kHz, and motion data is represented using SMPLX \cite{SMPL-X:2019} pose and FLAME \cite{FLAME:SiggraphAsia2017} expression parameters at 30 FPS. The text scripts are sourced from BEAT2 and formatted in MFA's TextGrid format.
  
\begin{table}[t]
\centering
\caption{\textbf{Quantitative evaluation on BEAT2.} We report FGD \(\times 10^{-1}\), BC \(\times 10^{-1}\), Diversity for body gestures, MSE \(\times 10^{-7}\), and LVD \(\times 10^{-5}\) for face gestures, UTMOS for speech audio, and average inference speed to generate per second gesture and audio (on NVIDIA 3090) to measure complexity. FastTalker significantly improves the inference speed, outperforms the previous state-of-the-art gesture generation methods, and performs on par with FastSpeech2.}
\resizebox{0.80\linewidth}{!}{
\begin{tabular}{lccccccc}
\toprule
                 & FGD $\downarrow$ & BC $\uparrow$ & Diversity~$\uparrow$ & MSE ~$\downarrow$ & LVD ~$\downarrow$ & UTMOS ~$\uparrow$ & Speed ~$\downarrow$ \\ 
\midrule
FaceFormer\cite{faceformer2022}       & -                      & -                    & -                    &  1.195                       & 8.824                       & -                       & -                         \\
CodeTalker\cite{xing2023codetalker}       & -                      & -                    & -                    & 1.243                  & 8.877                         & -                       & -                       \\
S2G\cite{ginosar2019learning}              & 25.129                 & \textbf{6.902}       & 7.783                & -                       & -                       & -                       & -                        \\
Trimodal\cite{yoon2020speech}         & 19.759                 & 6.442                & 8.894                & -                       & -                        & -                       & -                       \\
HA2G\cite{ha2g:liu2022learning}             & 19.364                 & 6.601                & 9.671                & -                       & -                       & -                       & -                        \\
DisCo\cite{liu2022disco}            & 21.170                 & 6.571                & 10.378               & -                       & -                       & -                       & -                        \\
CaMN\cite{liu2022beat}             & 8.752                  & 6.731                & 9.279                & -                       & -                        & -                       & -                       \\
DiffStyleGesture\cite{yang2023diffusestylegesture} & 10.137                  & 6.891                & 11.075                & -                       & -                      & -                & -                       \\
Habibie \textit{et al}.\cite{habibie2021learning}   & 14.581                 & 6.779                & 8.874                & 1.445                   & 9.548             & -                       & -                    \\
TalkShow \cite{talkshow:yi2022generating}         & 7.313                  & 6.783                & 12.859                & 1.399                   & 9.488            & -                       & -                    \\
EMAGE \cite{liu2023emage}     & 5.423                  & 6.794                & 13.057              & 1.180                   & 8.715                 & -                       & -                 \\      
\midrule
VITS \cite{kong2023vits2}   & -                  & -                & -              & -                 & -                & 3.77                       & - \\
FastSpeech \cite{ren2019fastspeech} & -                  & -                & -              & -                 & -                & 3.64                      & - \\
FastSpeech2 \cite{ren2020fastspeech} & -                  & -                & -              & -                 & -                & 3.91                       & - \\
\midrule
FastSpeech2+EMAGE & 5.413                  & 7.011                & 12.188              & 1.062                 & 8.491                & 3.91                      & 1.87 \\
FastTalker (Ours) & \textbf{4.891}                  & \textbf{7.613}                & \textbf{13.152 }            & \textbf{1.039}                 & \textbf{8.211}               & \textbf{3.93}                       & \textbf{0.17} \\
\bottomrule
\end{tabular}}
\label{tab:1}
\end{table}

\subsubsection{Metrics.} Following NaturalSpeech \cite{ju2024naturalspeech} and FastSpeech \cite{ren2020fastspeech}, we employ CMOS \cite{loizou2011speech} and UTMOS \cite{saeki2022utmos} to assess the quality of generated speech. The MOS (Mean Opinion Score) involves subjective ratings from 0 to 5 by invited judges for each speech audio group. The CMOS (Comparative Mean Opinion Score) compares these ratings between two result groups. The UTMOS (UTokyo-SaruLab Mean Opinion Score) is an objective metric derived from a pretrained network that predicts MOS values for speech samples. For gesture evaluation, we use FGD \cite{yoon2020speech}, Diversity \cite{li2021audio2gestures}, and BC \cite{li2021ai}, following the BEAT2 methodology \cite{liu2023emage}. Specifically, Featural Gesture Distance (FGD) measures the distribution distance between features of generated and ground truth gestures, extracted via pretrained SKCNN \cite{aberman2020skeleton}. Diversity calculates the average L1 distance across multiple gesture clips, and Beat Consistency (BC) compares the synchronization of audio beats with gesture motion beats. Additionally, we employ vertex MSE (Mean Square Error) \cite{xing2023codetalker} and vertex L1 Velocity Distance (LVD) \cite{talkshow:yi2022generating} to evaluate the positional and motion accuracies of generated facial animations.

\subsubsection{Training Settings.} Following the BEAT2 settings, we split the training data into training, validation, and testing sets with ratios of 8:1:1. Initially, we pretrain four VQVAEs for gesture and facial expression generation using rotation and vertex loss, subsequently fixing their weights. Optimization is performed on a single GPU with a learning rate of $3e-4$ using the ADAM optimizer \cite{kingma2014adam}. For the architecture search of modules in FastTalker, we adopt a two-layer LSTM as the controller, trained with an ADAM optimizer at a learning rate of $1e-3$. The search process utilizes only the speaker-2 data (4 hours), with the average training time for a candidate network being 1.8 hours. This entire search is executed over five days on 8 NVIDIA A100 40GB GPUs.

\subsection{Comparison with State-of-the-art}
We compare our method with state-of-the-art TTS and gesture generation methods including VITS \cite{kong2023vits2}, FastSpeech2 \cite{ren2020fastspeech} for speech synthesis, and EMAGE \cite{liu2023emage}, Talkshow \cite{yi2023generating}, Trimodal \cite{yoon2020speech}, FaceFormer \cite{faceformer2022}, and CodeTalker \cite{xing2023codetalker} for gesture generation. In addition, we compare our gesture generation method with a vanilla two-stage pipeline where speech synthesized by FastSpeech2 serves as input to EMAGE for generating gestures.

\begin{figure}[tb]
  \centering
  \includegraphics[width=0.8\textwidth]{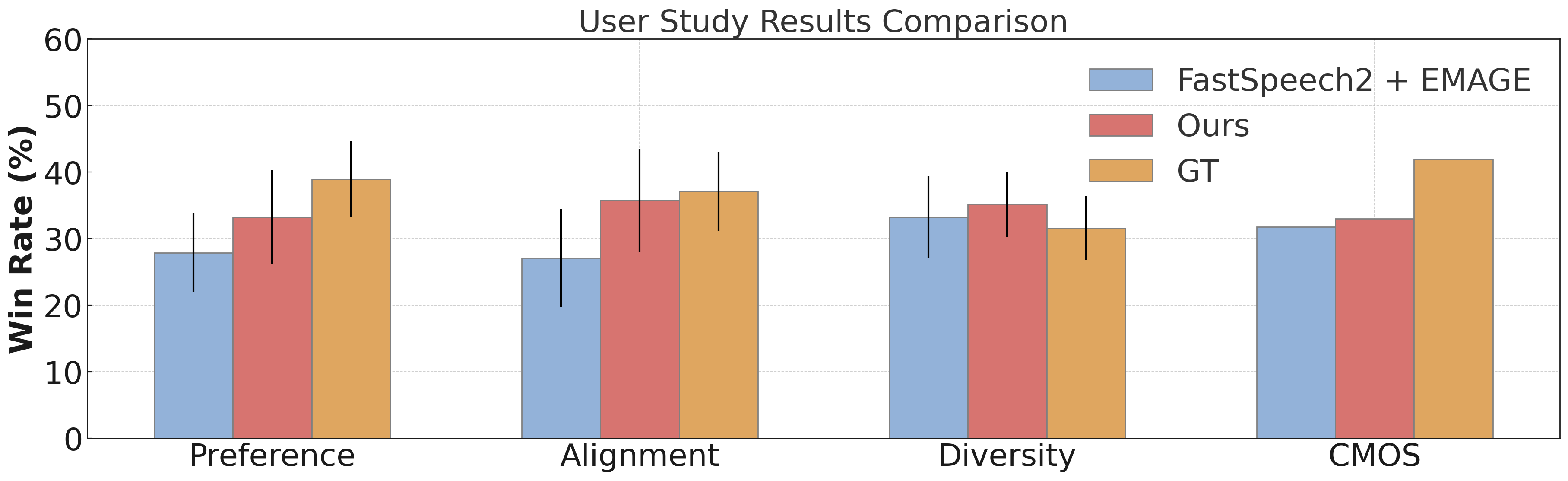}
\caption{\textbf{User Study Win Rate and CMOS.} Compared with ground truth, FastSpeech2 \cite{ren2020fastspeech} + EMAGE \cite{liu2023emage}, our FastTalker outperforms the separated FastSpeech2 + EMAGE stably with a higher win rate on gesture preference, audio-motion alignment, and gesture diversity. Additionally, the CMOS also demonstrates that FastTalker generates audio samples with similar quality to FastSpeech2.}

  \label{fig:4}
\end{figure}

\subsubsection{Objective evaluation on BEAT2.}
As shown in Table \ref{tab:1}, our method outperforms the previous state-of-the-art gesture generation method and performs on par with FastSpeech2. Additionally, it is more than 10 times faster with an average inference speed of 0.17s for generating per second speech and gesture on an NVIDIA 3090 GPU. Most importantly, the inference time being less than one second shows the potential for developing real-time applications.

\subsubsection{User study.}
\begin{figure}[tb]
  \centering
  \includegraphics[height=8cm]{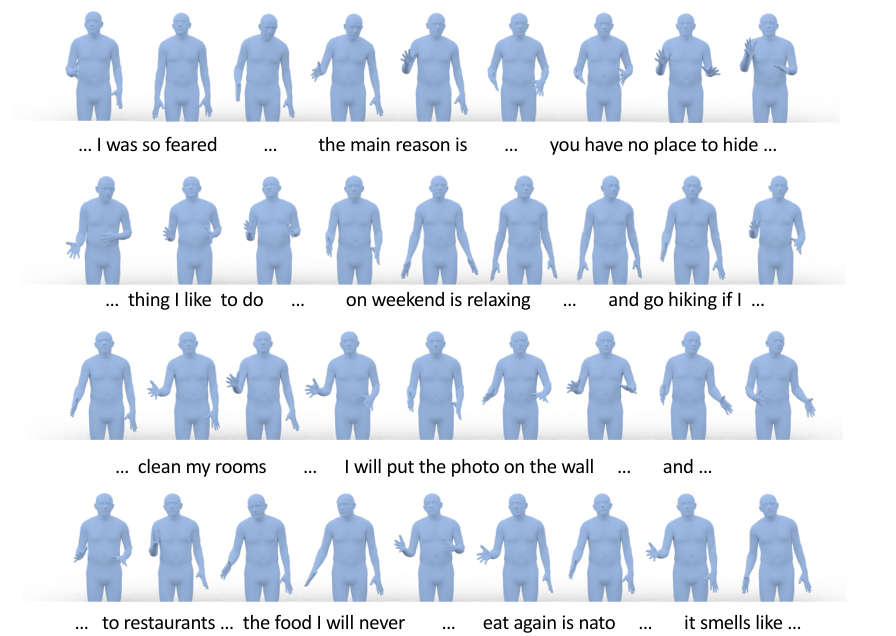}
\caption{\textbf{Subjective Generated Gesture Samples.} FastTalker not only generates results with a high audio-gesture rhythm alignment, but also semantically aligns with input text sentences. For example, it presents a ``\textit{reject}'' hand gesture when talking about the sentence ``\textit{never}''.}

  \label{fig:5}
\end{figure}

We invited 20 native speakers to make quality judgments about the synthesized speech and gesture samples on Google Forms. We sampled 120 comparison pairs between ground truth, FastSpeech2 + EMAGE, and our FastTalker, each with an equal duration ranging from 4 to 10 seconds. In a perceptual study, each participant evaluates a random set of 20 pairs during a 20-minute session by selecting the sequence they consider to have the best gesture preference, audio-gesture alignment, and gesture diversity. Additionally, we randomly sampled 10 generated speech samples from each method, with a total of 30 samples, and asked speakers to give the MOS score to the shuffled samples during another 10-minute session for CMOS. The results are shown in Fig. \ref{fig:4}. Furthermore, we visualize subjective generated results rendered with SMPLX \cite{SMPL-X:2019} in Fig. \ref{fig:5}. The results demonstrate that FastTalker also generates impressive gestures aligned with the semantics of the text sentence.

\subsection{Ablation}
\subsubsection{Baseline of unified framework with causal attention.} As shown in Table \ref{tab:2}, we start from the straightforward implementation with separated rhythmic predictors, where the refined phoneme features are fed into another group of rhythmic predictors for gesture generation. Using shared phoneme features results in a slight drop in performance. After replacing full self-attention with causal attention, the performance clearly drops in all terms, but it may be necessary for future real-time applications. 

\subsubsection{Effectiveness of Shared Rhythmic Predictors.} We further improve the speed of FastTalker by using shared rhythmic predictors. Feature sharing at the refined rhythm feature level, instead of the phoneme level, results in a clear improvement in audio-gesture alignment. Additionally, combining this with feature fusion, where the refined gesture-specific latent features are informed by reference speech duration, pitch, and energy features, can improve the fidelity of generated gestures without a clear performance drop in speech synthesis.

\vspace{-0.3cm}

\subsubsection{Effectiveness of neural architecture search.} The NAS significantly boosts the performance of FastTalker by optimizing the best hyperparameters for the phoneme encoder, rhythm predictors, semantic encoder, audio decoder, and gesture latent decoder. Surprisingly, we found the best number of layers for the gesture decoder and semantic encoder is zero, which means simple MLP decoders are enough to summarize semantic features and decode gestures, respectively. The final searched architecture result contains \({8, 4, 2, 4, 0, 8, 0}\) layers, \({256, 32, 64, 32, 64, 128, 128}\) channels, and \({4, 1, 1, 1, 8, 1, 4}\) group convolutions for the above seven modules in FastTalker. The result suggests that a computation-heavy phoneme embedding encoder is most important for the final performance and that shallow rhythm predictors are enough to regress the intermediate rhythm features such as duration, pitch, and energy.

\begin{table}[t]
\centering
\caption{\textbf{Ablation study results on BEAT2.}}
\resizebox{0.80\linewidth}{!}{
\begin{tabular}{lccccccc}
\toprule
               & FGD $\downarrow$ & BC~$\uparrow$ & Diversity $\uparrow$ & MSE~$\downarrow$ & LVD~$\downarrow$ & UTMOS~$\uparrow$ & Speed~$\downarrow$ \\ 
\midrule
Ground Truth   & 0                & 6.896             & 13.074               & 0                & 0                 & 4.28             & 0                 \\
Baseline       & 5.690            & 6.923             & 12.815               & 1.097            & 8.733             & 3.83             & 0.69              \\
+ Causal Block & 6.104            & 7.039             & 11.463               & 1.360            & 9.105             & 3.67             & 0.62             \\
+ Feature Fusion & 5.977          & 7.603             & 12.323               & 1.107            & 8.409             & 3.65             & 0.63            \\
+ NAS            & \textbf{4.891}       & \textbf{7.613}            & \textbf{13.152}               & \textbf{1.039}            & \textbf{8.211}             & \textbf{3.93}             & \textbf{0.17}   \\ 
\bottomrule
\end{tabular}}
\label{tab:2}
\end{table}
\vspace{-0.3cm}
\subsection{Limitation}
One limitation of FastTalker is the control of the speaker's voice; currently, our method can only generate timbres present in the training dataset. Integrating a voice conversion method to enable flexible timbre control may significantly increase the inference speed of our system. In particular, FastTalker generates speech and gestures in less than 0.2 seconds, but the existing state-of-the-art VC model \cite{qian2019autovc} takes 0.45 seconds per second to convert the voice to a target speaker. Including voice conversion in a unified framework will be future work.

\section{Conclusion}
We explore the task of jointly generating speech and full-body gestures from text scripts through our unified framework, FastTalker. The framework enhances speech-gesture alignment and inference efficiency by reusing intermediate rhythmic features. FastTalker is particularly suited for real-time applications such as talking avatars in live streaming, where online computation is crucial.

\bibliographystyle{splncs04}
\bibliography{main}

\begin{thebibliography}{10}
\providecommand{\url}[1]{\texttt{#1}}
\providecommand{\urlprefix}{URL }
\providecommand{\doi}[1]{https://doi.org/#1}

\bibitem{aberman2020skeleton}
Aberman, K., Li, P., Lischinski, D., Sorkine-Hornung, O., Cohen-Or, D., Chen,
  B.: Skeleton-aware networks for deep motion retargeting. ACM Transactions on
  Graphics (TOG)  \textbf{39}(4),  62--1 (2020)

\bibitem{ahuja2020style}
Ahuja, C., Lee, D.W., Nakano, Y.I., Morency, L.P.: Style transfer for co-speech
  gesture animation: A multi-speaker conditional-mixture approach. In: European
  Conference on Computer Vision. pp. 248--265. Springer (2020)

\bibitem{bhattacharya2021text2gestures}
Bhattacharya, U., Rewkowski, N., Banerjee, A., Guhan, P., Bera, A., Manocha,
  D.: Text2gestures: A transformer-based network for generating emotive body
  gestures for virtual agents. In: 2021 IEEE virtual reality and 3D user
  interfaces (VR). pp. 1--10. IEEE (2021)

\bibitem{brown2020language}
Brown, T., Mann, B., Ryder, N., Subbiah, M., Kaplan, J.D., Dhariwal, P.,
  Neelakantan, A., Shyam, P., Sastry, G., Askell, A., et~al.: Language models
  are few-shot learners. Advances in neural information processing systems
  \textbf{33},  1877--1901 (2020)

\bibitem{chen2020stabilizing}
Chen, X., Hsieh, C.J.: Stabilizing differentiable architecture search via
  perturbation-based regularization. In: International conference on machine
  learning. pp. 1554--1565. PMLR (2020)

\bibitem{cudeiro2019capture}
Cudeiro, D., Bolkart, T., Laidlaw, C., Ranjan, A., Black, M.J.: Capture,
  learning, and synthesis of 3d speaking styles. In: Proceedings of the
  IEEE/CVF conference on computer vision and pattern recognition. pp.
  10101--10111 (2019)

\bibitem{danvevcek2023emotional}
Dan{\v{e}}{\v{c}}ek, R., Chhatre, K., Tripathi, S., Wen, Y., Black, M.,
  Bolkart, T.: Emotional speech-driven animation with content-emotion
  disentanglement. In: SIGGRAPH Asia 2023 Conference Papers. pp. 1--13 (2023)

\bibitem{devlin2018bert}
Devlin, J., Chang, M.W., Lee, K., Toutanova, K.: Bert: Pre-training of deep
  bidirectional transformers for language understanding. arXiv preprint
  arXiv:1810.04805  (2018)

\bibitem{dong2024automated}
Dong, X., Kedziora, D.J., Musial, K., Gabrys, B., et~al.: Automated deep
  learning: Neural architecture search is not the end. Foundations and
  Trends{\textregistered} in Machine Learning  \textbf{17}(5),  767--920 (2024)

\bibitem{fan2022faceformer}
Fan, Y., Lin, Z., Saito, J., Wang, W., Komura, T.: Faceformer: Speech-driven 3d
  facial animation with transformers. In: Proceedings of the IEEE/CVF
  Conference on Computer Vision and Pattern Recognition. pp. 18770--18780
  (2022)

\bibitem{faceformer2022}
Fan, Y., Lin, Z., Saito, J., Wang, W., Komura, T.: Faceformer: Speech-driven 3d
  facial animation with transformers. In: Proceedings of the IEEE/CVF
  Conference on Computer Vision and Pattern Recognition (CVPR) (2022)

\bibitem{ginosar2019learning}
Ginosar, S., Bar, A., Kohavi, G., Chan, C., Owens, A., Malik, J.: Learning
  individual styles of conversational gesture. In: Proceedings of the IEEE/CVF
  Conference on Computer Vision and Pattern Recognition. pp. 3497--3506 (2019)

\bibitem{habibie2021learning}
Habibie, I., Xu, W., Mehta, D., Liu, L., Seidel, H.P., Pons-Moll, G., Elgharib,
  M., Theobalt, C.: Learning speech-driven 3d conversational gestures from
  video. arXiv preprint arXiv:2102.06837  (2021)

\bibitem{hua2024finematch}
Hua, H., Shi, J., Kafle, K., Jenni, S., Zhang, D., Collomosse, J., Cohen, S.,
  Luo, J.: Finematch: Aspect-based fine-grained image and text mismatch
  detection and correction. arXiv preprint arXiv:2404.14715  (2024)

\bibitem{hua2024v2xum}
Hua, H., Tang, Y., Xu, C., Luo, J.: V2xum-llm: Cross-modal video summarization
  with temporal prompt instruction tuning. arXiv preprint arXiv:2404.12353
  (2024)

\bibitem{huang2019voice}
Huang, W.C., Hayashi, T., Wu, Y.C., Kameoka, H., Toda, T.: Voice transformer
  network: Sequence-to-sequence voice conversion using transformer with
  text-to-speech pretraining. arXiv preprint arXiv:1912.06813  (2019)

\bibitem{jia2018transfer}
Jia, Y., Zhang, Y., Weiss, R., Wang, Q., Shen, J., Ren, F., Nguyen, P., Pang,
  R., Lopez~Moreno, I., Wu, Y., et~al.: Transfer learning from speaker
  verification to multispeaker text-to-speech synthesis. Advances in neural
  information processing systems  \textbf{31} (2018)

\bibitem{ju2024naturalspeech}
Ju, Z., Wang, Y., Shen, K., Tan, X., Xin, D., Yang, D., Liu, Y., Leng, Y.,
  Song, K., Tang, S., et~al.: Naturalspeech 3: Zero-shot speech synthesis with
  factorized codec and diffusion models. arXiv preprint arXiv:2403.03100
  (2024)

\bibitem{kim2021conditional}
Kim, J., Kong, J., Son, J.: Conditional variational autoencoder with
  adversarial learning for end-to-end text-to-speech. In: International
  Conference on Machine Learning. pp. 5530--5540. PMLR (2021)

\bibitem{kingma2014adam}
Kingma, D.P., Ba, J.: Adam: A method for stochastic optimization. arXiv
  preprint arXiv:1412.6980  (2014)

\bibitem{kong2023vits2}
Kong, J., Park, J., Kim, B., Kim, J., Kong, D., Kim, S.: Vits2: Improving
  quality and efficiency of single-stage text-to-speech with adversarial
  learning and architecture design. arXiv preprint arXiv:2307.16430  (2023)

\bibitem{li2023zico}
Li, G., Yang, Y., Bhardwaj, K., Marculescu, R.: Zico: Zero-shot nas via inverse
  coefficient of variation on gradients. arXiv preprint arXiv:2301.11300
  (2023)

\bibitem{li2021audio2gestures}
Li, J., Kang, D., Pei, W., Zhe, X., Zhang, Y., He, Z., Bao, L.: Audio2gestures:
  Generating diverse gestures from speech audio with conditional variational
  autoencoders. In: Proceedings of the IEEE/CVF International Conference on
  Computer Vision. pp. 11293--11302 (2021)

\bibitem{li2019neural}
Li, N., Liu, S., Liu, Y., Zhao, S., Liu, M.: Neural speech synthesis with
  transformer network. In: Proceedings of the AAAI conference on artificial
  intelligence. vol.~33, pp. 6706--6713 (2019)

\bibitem{li2021ai}
Li, R., Yang, S., Ross, D.A., Kanazawa, A.: Ai choreographer: Music conditioned
  3d dance generation with aist++. In: Proceedings of the IEEE/CVF
  International Conference on Computer Vision. pp. 13401--13412 (2021)

\bibitem{FLAME:SiggraphAsia2017}
Li, T., Bolkart, T., Black, M.J., Li, H., Romero, J.: Learning a model of
  facial shape and expression from {4D} scans. ACM Transactions on Graphics,
  (Proc. SIGGRAPH Asia)  \textbf{36}(6),  194:1--194:17 (2017),
  \url{https://doi.org/10.1145/3130800.3130813}

\bibitem{li2024feature}
Li, Z., Huang, Y., Zhu, M., Zhang, J., Chang, J., Liu, H.: Feature manipulation
  for ddpm based change detection. arXiv preprint arXiv:2403.15943  (2024)

\bibitem{li2023stock}
Li, Z., Yu, H., Xu, J., Liu, J., Mo, Y.: Stock market analysis and prediction
  using lstm: A case study on technology stocks. Innovations in Applied
  Engineering and Technology pp.~1--6 (2023)

\bibitem{liu2022disco}
Liu, H., Iwamoto, N., Zhu, Z., Li, Z., Zhou, Y., Bozkurt, E., Zheng, B.: Disco:
  Disentangled implicit content and rhythm learning for diverse co-speech
  gestures synthesis. In: Proceedings of the 30th ACM International Conference
  on Multimedia. pp. 3764--3773 (2022)

\bibitem{liu2020reinforcement}
Liu, H., Zhang, C.: Reinforcement learning based neural architecture search for
  audio tagging. In: 2020 International Joint Conference on Neural Networks
  (IJCNN). pp.~1--8. IEEE (2020)

\bibitem{liu2023emage}
Liu, H., Zhu, Z., Becherini, G., Peng, Y., Su, M., Zhou, Y., Iwamoto, N.,
  Zheng, B., Black, M.J.: Emage: Towards unified holistic co-speech gesture
  generation via masked audio gesture modeling. arXiv preprint arXiv:2401.00374
   (2023)

\bibitem{liu2022beat}
Liu, H., Zhu, Z., Iwamoto, N., Peng, Y., Li, Z., Zhou, Y., Bozkurt, E., Zheng,
  B.: Beat: A large-scale semantic and emotional multi-modal dataset for
  conversational gestures synthesis. arXiv preprint arXiv:2203.05297  (2022)

\bibitem{liu2018darts}
Liu, H., Simonyan, K., Yang, Y.: Darts: Differentiable architecture search.
  arXiv preprint arXiv:1806.09055  (2018)

\bibitem{liu2022learning}
Liu, X., Wu, Q., Zhou, H., Xu, Y., Qian, R., Lin, X., Zhou, X., Wu, W., Dai,
  B., Zhou, B.: Learning hierarchical cross-modal association for co-speech
  gesture generation. In: Proceedings of the IEEE/CVF Conference on Computer
  Vision and Pattern Recognition. pp. 10462--10472 (2022)

\bibitem{ha2g:liu2022learning}
Liu, X., Wu, Q., Zhou, H., Xu, Y., Qian, R., Lin, X., Zhou, X., Wu, W., Dai,
  B., Zhou, B.: Learning hierarchical cross-modal association for co-speech
  gesture generation. In: Proceedings of the IEEE/CVF Conference on Computer
  Vision and Pattern Recognition. pp. 10462--10472 (2022)

\bibitem{loizou2011speech}
Loizou, P.C.: Speech quality assessment. In: Multimedia analysis, processing
  and communications, pp. 623--654. Springer (2011)

\bibitem{lu2023artistic}
Lu, Q., Lee, J., Endo, Y., Kamijo, S.: Artistic line drawing rendering with
  priors of depth and edge density. IEEE MultiMedia  (2023)

\bibitem{lyu2022study}
Lyu, W., Zheng, S., Ma, T., Chen, C.: A study of the attentionabnormality in
  trojaned berts. In: Proceedings of the2022 Conference of the North American
  Chapter of the Association for Computational Linguistics: Human Language
  Technologies. pp. 4727--4741 (2022)

\bibitem{lyu2023attention}
Lyu, W., Zheng, S., Pang, L., Ling, H., Chen, C.: Attention-enhancing backdoor
  attacks against bert-based models. In: Findings of the Association for
  Computational Linguistics: EMNLP 2023. pp. 10672--10690 (2023)

\bibitem{pang2023bodyformer}
Pang, K., Qin, D., Fan, Y., Habekost, J., Shiratori, T., Yamagishi, J., Komura,
  T.: Bodyformer: Semantics-guided 3d body gesture synthesis with transformer.
  ACM Transactions on Graphics (TOG)  \textbf{42}(4),  1--12 (2023)

\bibitem{SMPL-X:2019}
Pavlakos, G., Choutas, V., Ghorbani, N., Bolkart, T., Osman, A.A.A., Tzionas,
  D., Black, M.J.: Expressive body capture: {3D} hands, face, and body from a
  single image. In: Proceedings IEEE Conf. on Computer Vision and Pattern
  Recognition (CVPR). pp. 10975--10985 (2019)

\bibitem{peng2020non}
Peng, K., Ping, W., Song, Z., Zhao, K.: Non-autoregressive neural
  text-to-speech. In: International conference on machine learning. pp.
  7586--7598. PMLR (2020)

\bibitem{peng2023emotalk}
Peng, Z., Wu, H., Song, Z., Xu, H., Zhu, X., He, J., Liu, H., Fan, Z.: Emotalk:
  Speech-driven emotional disentanglement for 3d face animation. In:
  Proceedings of the IEEE/CVF International Conference on Computer Vision. pp.
  20687--20697 (2023)

\bibitem{qian2019autovc}
Qian, K., Zhang, Y., Chang, S., Yang, X., Hasegawa-Johnson, M.: Autovc:
  Zero-shot voice style transfer with only autoencoder loss. In: International
  Conference on Machine Learning. pp. 5210--5219. PMLR (2019)

\bibitem{ren2020fastspeech}
Ren, Y., Hu, C., Tan, X., Qin, T., Zhao, S., Zhao, Z., Liu, T.Y.: Fastspeech 2:
  Fast and high-quality end-to-end text to speech. arXiv preprint
  arXiv:2006.04558  (2020)

\bibitem{ren2019fastspeech}
Ren, Y., Ruan, Y., Tan, X., Qin, T., Zhao, S., Zhao, Z., Liu, T.Y.: Fastspeech:
  Fast, robust and controllable text to speech. Advances in neural information
  processing systems  \textbf{32} (2019)

\bibitem{richard2021meshtalk}
Richard, A., Zollh{\"o}fer, M., Wen, Y., De~la Torre, F., Sheikh, Y.: Meshtalk:
  3d face animation from speech using cross-modality disentanglement. In:
  Proceedings of the IEEE/CVF International Conference on Computer Vision. pp.
  1173--1182 (2021)

\bibitem{saeki2022utmos}
Saeki, T., Xin, D., Nakata, W., Koriyama, T., Takamichi, S., Saruwatari, H.:
  Utmos: Utokyo-sarulab system for voicemos challenge 2022. arXiv preprint
  arXiv:2204.02152  (2022)

\bibitem{shen2023naturalspeech}
Shen, K., Ju, Z., Tan, X., Liu, Y., Leng, Y., He, L., Qin, T., Zhao, S., Bian,
  J.: Naturalspeech 2: Latent diffusion models are natural and zero-shot speech
  and singing synthesizers. arXiv preprint arXiv:2304.09116  (2023)

\bibitem{sutton2000policy}
Sutton, R.S., McAllester, D., Singh, S., Mansour, Y.: Policy gradient methods
  for reinforcement learning with function approximation. Advances in neural
  information processing systems  \textbf{12} (2000)

\bibitem{tan2024naturalspeech}
Tan, X., Chen, J., Liu, H., Cong, J., Zhang, C., Liu, Y., Wang, X., Leng, Y.,
  Yi, Y., He, L., et~al.: Naturalspeech: End-to-end text-to-speech synthesis
  with human-level quality. IEEE Transactions on Pattern Analysis and Machine
  Intelligence  (2024)

\bibitem{tang2024avicuna}
Tang, Y., Shimada, D., Bi, J., Xu, C.: Avicuna: Audio-visual llm with
  interleaver and context-boundary alignment for temporal referential dialogue.
  arXiv preprint arXiv:2403.16276  (2024)

\bibitem{xie2021weight}
Xie, L., Chen, X., Bi, K., Wei, L., Xu, Y., Wang, L., Chen, Z., Xiao, A.,
  Chang, J., Zhang, X., et~al.: Weight-sharing neural architecture search: A
  battle to shrink the optimization gap. ACM Computing Surveys (CSUR)
  \textbf{54}(9),  1--37 (2021)

\bibitem{xing2023codetalker}
Xing, J., Xia, M., Zhang, Y., Cun, X., Wang, J., Wong, T.T.: Codetalker:
  Speech-driven 3d facial animation with discrete motion prior. In: Proceedings
  of the IEEE/CVF Conference on Computer Vision and Pattern Recognition. pp.
  12780--12790 (2023)

\bibitem{yang2023diffusestylegesture}
Yang, S., Wu, Z., Li, M., Zhang, Z., Hao, L., Bao, W., Cheng, M., Xiao, L.:
  Diffusestylegesture: Stylized audio-driven co-speech gesture generation with
  diffusion models. arXiv preprint arXiv:2305.04919  (2023)

\bibitem{yang2021towards}
Yang, Y., You, S., Li, H., Wang, F., Qian, C., Lin, Z.: Towards improving the
  consistency, efficiency, and flexibility of differentiable neural
  architecture search. In: Proceedings of the IEEE/CVF conference on computer
  vision and pattern recognition. pp. 6667--6676 (2021)

\bibitem{ye2022b}
Ye, P., Li, B., Li, Y., Chen, T., Fan, J., Ouyang, W.: b-darts: Beta-decay
  regularization for differentiable architecture search. In: proceedings of the
  IEEE/CVF conference on computer vision and pattern recognition. pp.
  10874--10883 (2022)

\bibitem{yi2023generating}
Yi, H., Liang, H., Liu, Y., Cao, Q., Wen, Y., Bolkart, T., Tao, D., Black,
  M.J.: Generating holistic 3d human motion from speech. In: Proceedings of the
  IEEE/CVF Conference on Computer Vision and Pattern Recognition. pp. 469--480
  (2023)

\bibitem{talkshow:yi2022generating}
Yi, H., Liang, H., Liu, Y., Cao, Q., Wen, Y., Bolkart, T., Tao, D., Black,
  M.J.: Generating holistic 3d human motion from speech. In: CVPR (2023)

\bibitem{yoon2020speech}
Yoon, Y., Cha, B., Lee, J.H., Jang, M., Lee, J., Kim, J., Lee, G.: Speech
  gesture generation from the trimodal context of text, audio, and speaker
  identity. ACM Transactions on Graphics (TOG)  \textbf{39}(6),  1--16 (2020)

\bibitem{yu2022cyclic}
Yu, H., Peng, H., Huang, Y., Fu, J., Du, H., Wang, L., Ling, H.: Cyclic
  differentiable architecture search. IEEE transactions on pattern analysis and
  machine intelligence  \textbf{45}(1),  211--228 (2022)

\bibitem{yu2024promptfix}
Yu, Y., Zeng, Z., Hua, H., Fu, J., Luo, J.: Promptfix: You prompt and we fix
  the photo. arXiv preprint arXiv:2405.16785  (2024)

\bibitem{zhang2024research}
Zhang, J., Cao, J., Chang, J., Li, X., Liu, H., Li, Z.: Research on the
  application of computer vision based on deep learning in autonomous driving
  technology. arXiv preprint arXiv:2406.00490  (2024)

\end{thebibliography}
\end{document}